\documentclass[prl%
,twocolumn%
,floats
,aps%
,showpacs%
]{revtex4}
\usepackage{amsmath}%

\begin{document}
\title{Numerical Results for Ground States of Mean-Field Spin Glasses at low Connectivities} 
\date{\today}
\author{Stefan Boettcher}  
\email{www.physics.emory.edu/faculty/boettcher}
\affiliation{Physics Department, Emory University, Atlanta, Georgia
30322, USA}  

\begin{abstract} 
An extensive list of results for the ground state properties of spin
glasses on random graphs is presented. These results provide a timely
benchmark for currently developing theoretical techniques based on
replica symmetry breaking that are being tested on mean-field models
at low connectivity. Comparison with existing replica results for such
models verifies the strength of those techniques. Yet, we find that
spin glasses on fixed-connectivity graphs (Bethe lattices) exhibit a
richer phenomenology than has been anticipated by theory. Our data prove
to be sufficiently accurate to speculate about some exact results.
\end{abstract}  
\maketitle
A theoretical understanding of the intricate dynamics of disordered
systems has been a major goal of statistical physics at least since
the introduction of the Edwards-Anderson spin glass model
\cite{EA}. Already the study of the equilibrium at low temperature, a
state real disordered materials rarely achieve \cite{FH}, reveals a
stunning range of new phenomena, even in the simplest models such as
the Sherrington and Kirkpatrick model (SK) where all spins are
mutually connected \cite{SK,TAP,ParisiRSB,MPV}.  As an intermediate
step in extending the mean-field techniques toward finite-dimensional
models, spin glasses on random graphs are an area of active
research. Those systems have been of interest from early on because
they combine infinite-range connections (like SK) with a finite,
decidedly low connectivity. But those earlier studies have focused
either on temperatures at the glass transition \cite{VB}, on purely
replica symmetric (RS) solutions \cite{KaSo}, or on perturbative
approaches in the (SK-)limit of large connectivity
\cite{PT,DG,LG}. 

Simultaneously, the formal similarity between
spin-glass Hamiltonians and the objective function of combinatorial
optimization problems has been realized \cite{SA} and exploited to
make RS predictions, for instance, for the bipartitioning problem on
random graphs \cite{MPGB,WS,Banavar}. This connection, with the
discovery of phase transitions in combinatorial optimization problems
\cite{HH} and the application of replica techniques to their study
\cite{Monasson}, has recently rejuvenated interest in spin glasses on
random graphs. But to obtain quantitatively valuable predictions for
NP-hard problems required the application of replica symmetry breaking
(RSB) \cite{ParisiRSB} to those problems at finite connectivities and
low temperatures which was accomplished recently
\cite{MP1,Franz,MPZ}. Finally, these RSB methods are now being applied
to spin glasses on random graphs, producing quantitatively valuable
results \cite{MP1,MP2} at accuracies below 0.1\%. At this level of
accuracy, a comparison between theoretical and simulation results
becomes valuable at least in two respects: Convergence of the
numerical with the RSB result can verify the
assumptions underlying RSB as well as the quality of the numerical
method used to approximate an NP-hard problem.

In this letter we apply the extremal optimization (EO) heuristic
\cite{BoPe1,eo_prl} to investigate the ground state properties of spin
glasses on random graphs. With this method we have sampled system
sizes up to $n=4096$ on low-connectivity graphs. We have obtained
high-accuracy results for the ground-state energies of spin glasses on
ordinary random graphs (ORG) with fluctuating connectivities, and for
Bethe-lattice graphs (BL) with fixed connectivities. On a smaller
sample of BL, we have also obtained results for the entropy of such
graphs. The energies are in excellent agreement with RSB predictions
for low-connectivity BL \cite{MP2}. Both, the energies and entropies
reveal a sensitivity to the even-oddness of the BL, which may explain
inconsistencies with results in the SK limit \cite{PT}. No such
inconsistency arises for ORG and the numerical extrapolation is in
good agreement with analytic results for the large-connectivity limit
in RSB \cite{PT}. The BL energies seem to fall (within $0.2\%$) on a
simple line ranging from the 2-connected graphs to the SK-limit, but
only for {\it even integer} connectivities, without any obvious
interpolation. The BL entropies decrease linearly with the inverse
connectivity for odd connectivities, but are already consistent with
zero (within accuracy) at small even connectivity. More details of the
numerical procedure is given elsewhere \cite{eo_bethe,eo_rg}.

Of the two types of random graphs are considered for this study, the
BL are regular random graphs \cite{Bollobas}. These graphs consist of
$n$ vertices where each vertex possesses a fixed number $k+1$ of bonds
\cite{MPGB,MP1,MP2} with randomly selected other vertices.
Alternatively, ORG are obtained by randomly connecting any pair of
vertices with a specified probability $p=c/(n-1)$, leading to a graph
of average connectivity $c$ but where the connectivities of individual
vertices are Poissonian distributed \cite{Bollobas}. Note that each vertex' connectivity, and thus $k+1$, is inherently discrete, while $c$ can take on any real value.

Once a graph of connectivity $c$ is generated, randomly chosen
quenched couplings $J_{i,j}\in\{-1,+1\}$ are assigned to existing
bonds between neighboring vertices $i$ and $j$. Each vertex $i$ is
occupied by a spin variable $x_i\in\{-1,+1\}$. The energy of the
system is defined as the difference in number between violated bonds
and satisfied bonds, $H=-\sum_{\{bonds\}} J_{i,j} x_i x_j$, and we
will focus on the energy and entropy per spin, resp.,
\begin{eqnarray}
e_c=\frac{1}{n}\,H,\qquad s_c=\frac{1}{n}\,\ln\Omega,
\label{eeq}
\end{eqnarray}
where $\Omega$ is the degeneracy of the configurations exhibiting the
ground state energy.

For our numerical procedure we used the following implementation of EO
\cite{BoPe1,eo_prl}: For a given spin configuration on a graph, assign
to each spin $x_i$ a ``fitness''
$\lambda_i=-\#violated~bonds=-0,-1,-2,\ldots,-c_i$, so that
$e_c=-\sum_i\lambda_i/(2n)$ is satisfied. Here, $c_i$ is the integer
connectivity of vertex $i$, and $c_i\equiv k+1$ for every vertex in
BL. If $c_{\rm max}=\max_ic_i$, each spin falls into one of only
$c_{\rm max}+1$ possible states. Say, currently there are $n_{c_{\rm
max}}$ spins with the worst fitness, $\lambda=-c_{\rm max}$,
$n_{c_{\rm max}-1}$ with $\lambda=-c_{\rm max}+1$, and so on up to
$n_0$ spins with the best fitness $\lambda=0$, where
$\sum_jn_j=n$. Now draw a ``rank'' $l$ according to the distribution
$P(l)\sim l^{-\tau}$.  Then, determine $0\leq j\leq c_{\rm max}$ such
that $\sum_{i=j+1}^{c_{\rm max}}n_i<l\leq\sum_{i=j}^{c_{\rm
max}}n_i$. Finally, select any one of the $n_j$ spins in state $j$ and
reverse its orientation {\em unconditionally.} As a result, it and its
neighboring spins change their fitness. After all the effected
$\lambda$'s and $n$'s are reevaluated, a new spin is chosen for an
update.
 
The arguments given in \cite{eo_jam} and a few experiments indicate
that $\tau=1.3$ is a satisfactory choice to find ground states
efficiently on either type of graph.  Our implementation restarts for
each instance at least $r_{\rm max}=4$ times with new random initial
spin assignments, executing $\approx0.1\,n^3$ updates per run. If a
new, lower-than-previous energy state is encountered in run $r$, we
adjust $r_{\rm max}=2+2r$ for that instance so that EO runs at least
twice as many restarts as were necessary to find the lowest state in
the first place. Especially for small $n$, $r_{\rm max}$ hardly ever
exceeds 4; for larger $n$ a few graphs require up to 30 restarts
before termination.  Since EO perpetually explores new configurations
it is well suited to explore also the degeneracy $\Omega$ of
low-energy states. In these runs, we used a similar approach to the
above, except for setting $r_{\rm max}=8+2r$, where $r$ is the latest
run in which a new configuration of the lowest energy was located.

\begin{figure}
\vskip 4.2in 
\includegraphics{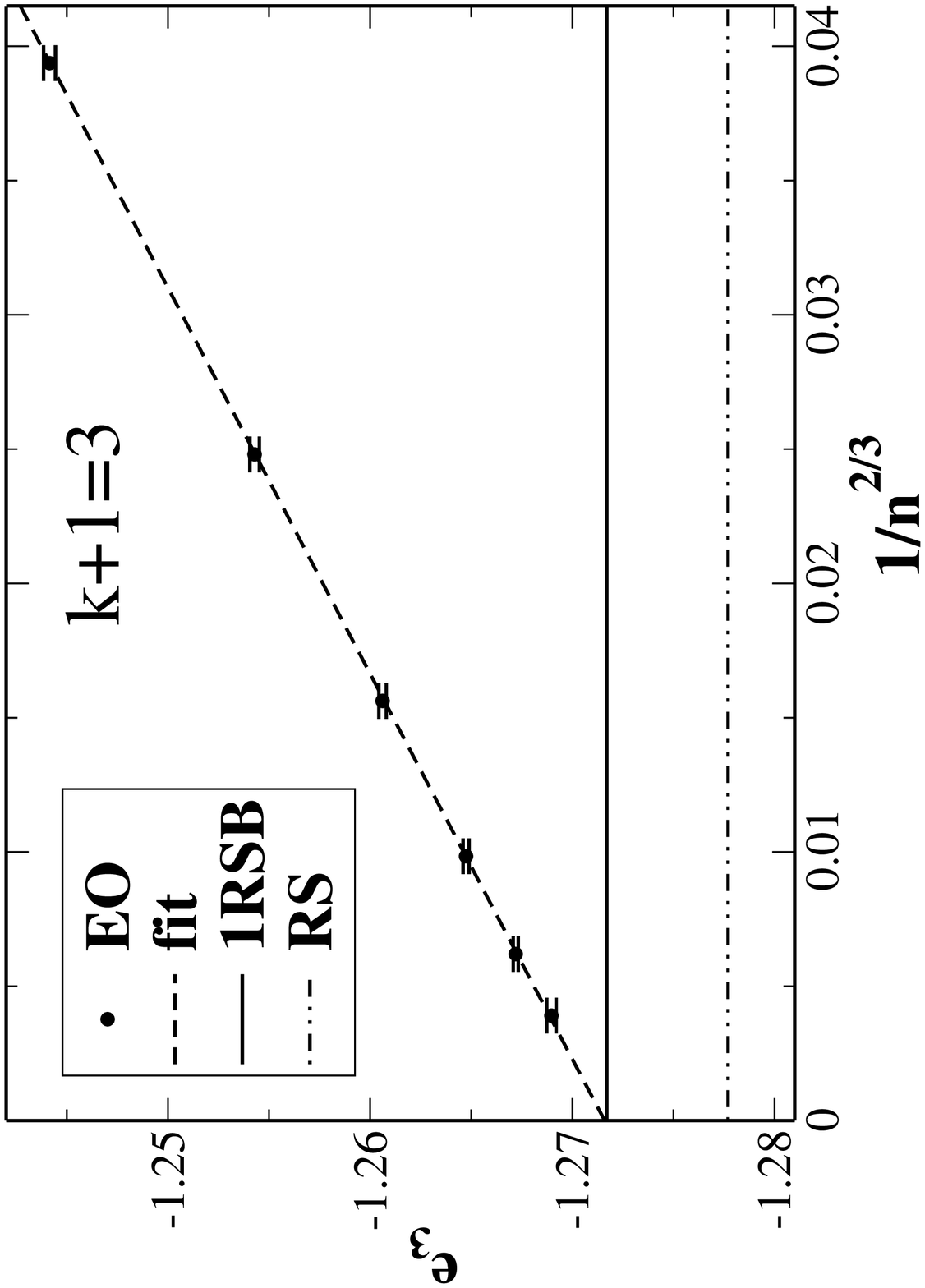}
\includegraphics{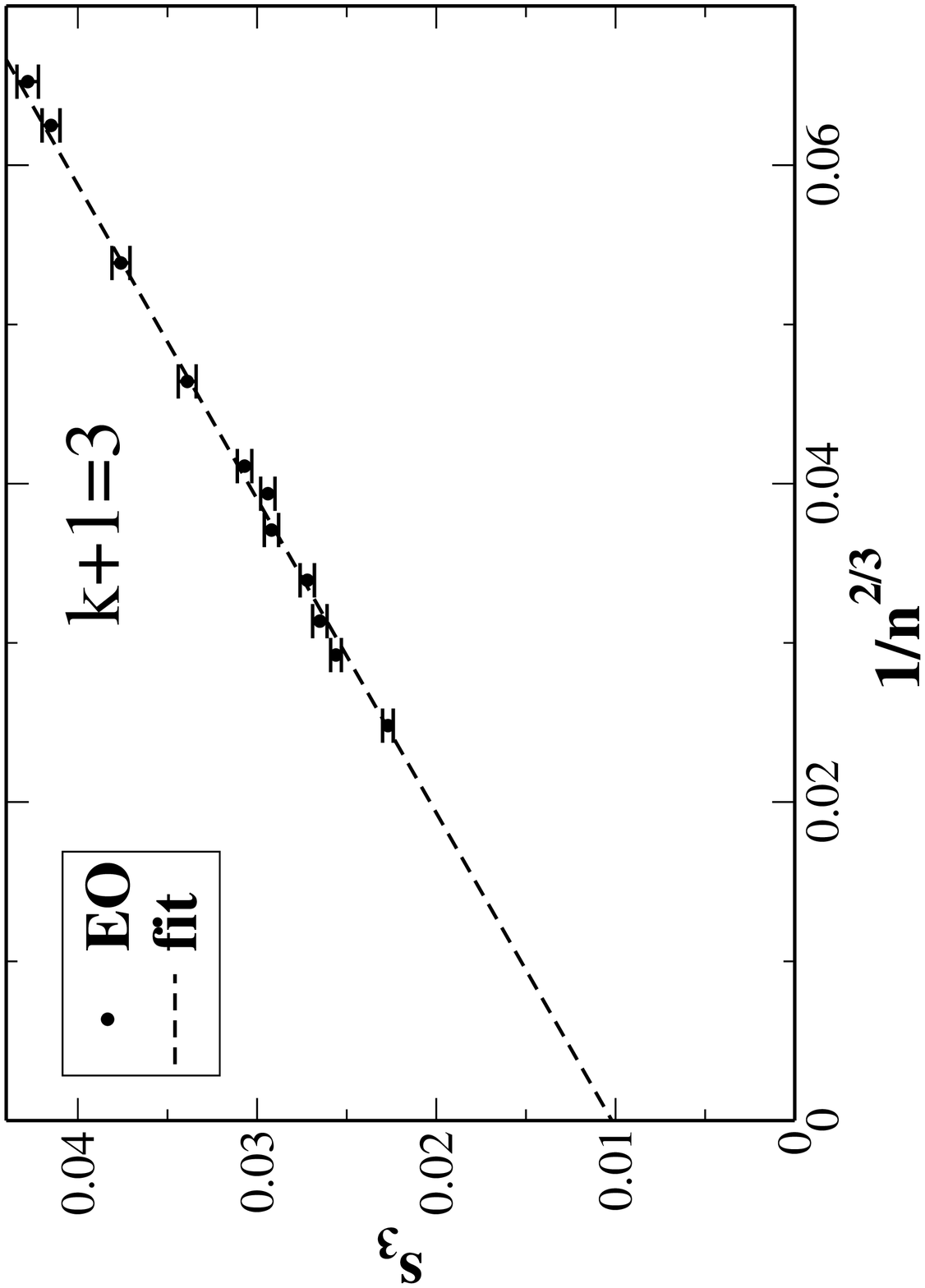}
\caption{Extrapolation plot for the BL energies (above) and entropies
(below) for $k+1=3$ obtained with EO for finite $n$ \protect\cite{eo_bethe}. The data is plotted vs. $1/n^{2/3}$ and is fitted according to
Eq.~(\protect\ref{extrapolationeq}).  (Data points were weighted with
respect to $n$ and the inverse of the error.)  For $n\to\infty$ the
extrapolation for the energy gives $e_3=-1.2716(1)$, way above the RS
result but consistent with the 1RSB result~\protect\cite{MP2}, both
indicated by horizontal lines. The extrapolation for the entropy gives
$s_3=0.0102(10)$. }
\label{bethe3plot}
\end{figure}

We have simulated spin glasses on BL with this algorithm for $k+1$
between 3 and 26, and graph sizes $n=2^l$ for $l=5,6,\ldots,10$ to
obtain results for ground state energies \cite{eo_bethe}. In
particular, for $k+1=3$ we have used the methods described in
Ref.~\cite{eo_rg} to reach system sizes of $n=4096$. In a separate
simulation, using $\tau=1.4$, we have explored BL of size
$n\in[16\ldots256]$ to determine their entropy. We have used the same
algorithm, preceded by a graph reduction procedure \cite{eo_rg}, to
study ORG ranging from $n\leq2^{15}$ for $c=2$ to $n\leq2^9$ at
$c=25$. Amazingly, as is shown in Refs.~\cite{eo_bethe,eo_rg}, in all
these cases our data can be extrapolated for $n\to\infty$ via
\begin{eqnarray}
e_c(n)\sim e_c+\frac{A}{n^{2/3}}\quad(n\to\infty).
\label{extrapolationeq}
\end{eqnarray}
Deviations from these scaling corrections are generally small
\cite{eo_bethe} and we assume Eq.~(\ref{extrapolationeq}) to be exact
here. There does not appear to be a theoretical justification for
Eq.~(\ref{extrapolationeq}), but in Fig.~\ref{bethe3plot} we show
representatively the large-$n$ extrapolation of our data for
$k+1=3$. Both, finite-$n$ energies and entropies, appear to scale
linearly when plotted for $1/n^{2/3}$. The extrapolation results for
the energies of BL and ORG are given in Tab.~\ref{enertable}. The
extrapolation results for the BL entropies are given in
Tab.~\ref{Betheentrotable}. Variation in the (estimated) errors
reflect differences in computational effort (number of instances,
largest $n$) and in the quality of the extrapolation \cite{eo_bethe}.

\begin{table}
\caption{Extrapolated energies per spin for BL (left) and ORG
(right). Although only integer values of the average connectivity
$c$ were considered, it can take on any real value, unlike $k+1$.}
\begin{tabular}{cl|cl||cl}
\hline\hline
$k+1$ &  $~~~e_{k+1}$ & $k+1$ &  $~~~e_{k+1}$ &  $c$ &  $~~~e_c$\\
\hline 
3  & -1.2716(1) & 12 & -2.6127(9) & 2 &      -0.9192(2)\\
4  & -1.472(1)  & 14 & -2.8287(5) & 3 &      -1.2059(2) \\
5  & -1.673(1)  & 15 & -2.935(1)  & 4 &      -1.4311(10) \\
6  & -1.826(1)  & 16 & -3.0268(9) & 5 &      -1.6224(10)\\
7  & -1.991(3)  & 18 & -3.212(2)  & 10 &  -2.356(3)\\
8  & -2.1213(9)  & 20 & -3.389(1) & 15 &  -2.906(5)\\
9  & -2.2645(5) & 25 & -3.806(4)  & 20 &  -3.373(5)\\
10 & -2.378(3)  &    &            & 25 &  -3.775(8)\\
\hline\hline
\end{tabular}
\label{enertable}
\end{table}
\begin{table}
\caption{Extrapolated entropies per spin for BL.}
\begin{tabular}{cl|cl}
\hline\hline
$k+1$ & $~~~s_{k+1}$ & $k+1$ &  $~~~s_{k+1}$\\
\hline
3&	0.0102(10)& 4&	0.0381(15)\\
5&	0.0048(10)& 6&	0.0291(10)\\
7&	0.0020(10)& 8&	0.0218(10)\\
9&	0.0002(15)& 10&	0.0198(10)\\
14&	0.0126(10)& 15& 0.0002(15)\\
18&	0.0095(10)& 22&	0.0076(10)\\
26&	0.0063(15)&&\\
\hline\hline
\end{tabular}
\label{Betheentrotable}
\end{table}

We can compare our results with existing theoretical predictions at
the RS and the 1RSB level at least for the case of BL at $k+1=3$. For
this case, Ref.~\cite{MP2} reproduced $e_3=-1.2777$ at the RS level,
and yielded $e_3=-1.2717$ at the 1RSB level (further replica
corrections are expected to be small). These values and our
extrapolation result of $e_3=-1.2716(1)$ are indicated in
Fig.~\ref{bethe3plot}. Clearly, the extrapolation result is extremely
close to the 1RSB results, but inconsistent with the RS result.

We have also used EO to sample the degeneracy $\Omega$ of the
lowest-energy states found for BL. In these simulations we focused on
smaller system sizes of $n\leq256$ for $k+1=3,\ldots,9$ and
$10,14,\ldots,26$ only. As a test for the accuracy of our
implementation, we have run the simulation for $k+1=3$ a second time
on identical instances, using different initial conditions and $n/5$
more updates, obtaining identical results {\it for each instance.}

\begin{figure}
\vskip 2.1in \includegraphics{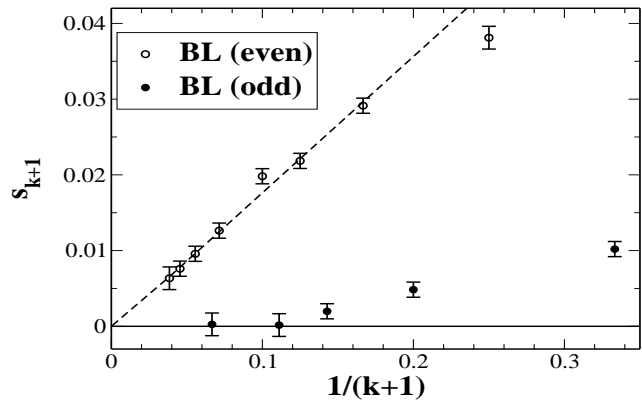}
\caption{Asymptotic plot of the extrapolated entropies from
Tab.~\protect\ref{Betheentrotable} as a function of $1/(k+1)$. The
data for even $k+1$ seems to vanish linearly with $1/(k+1)$ (dashed
line). The data for odd $k+1$ drops more precipitously, and can not
reasonably be fitted at this level of accuracy. }
\label{entroscalplot}
\end{figure}

When plotted for $1/(k+1)$ in Fig.~\ref{entroscalplot}, the BL
entropies for even connectivities decrease about linearly toward
zero. The entropies for odd connectivities, clearly non-zero at
$k+1=3$ (see Fig.~\ref{bethe3plot}), drop more rapidly and are
essentially indistinguishable from zero already at $k+1=9$.  While any
rapid, smooth decay could easily escape our limited accuracy, the plot
still raises the question whether there may be a finite connectivity
beyond which odd entropies vanish.  The qualitative difference between
even and odd BL entropies can be understood in the presents or
absence, resp., of ``free spins,'' a finite fraction of spins in the
ground state which violate exactly half of their bonds and may flip at
no cost \cite{eo_bethe}.

\begin{figure}
\vskip 2.2in \includegraphics{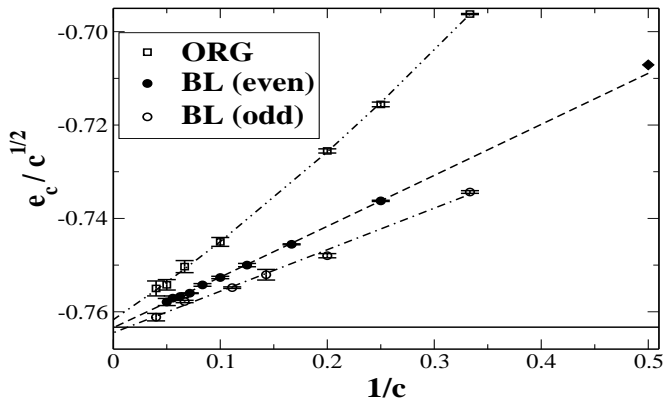}
\caption{Plot of the rescaled extrapolated energies, $e_{c}/\sqrt{c}$,
as a function of $1/c$ for ORG (squares) and BL (circles), where
$c=k+1$.  The BL data appears to fall on two separate straight lines
for even and for odd $k+1$, {\it including} the trivial result,
$e_2=-1$ (diamond). In all cases, the fits (dashed lines) provide an
reasonable estimate for $E_{SK}=-0.76321$ (horizontal line) at
infinite connectivity.  }
\label{SKplot}
\end{figure}

We have also plotted all BL and ORG energies as $e_c/\sqrt{c}$
vs. $1/c$ (where $c=k+1$ for BL) in Fig.~\ref{SKplot}. We expect that
$\lim_{c\to\infty}e_{c}/\sqrt{c}=E_{SK}=-0.76321$ for RSB
\cite{CR,MPV}. All energies for ORG appear to fall on a single smooth
curve. We can fit those energies with a parabola, projecting
$E_{SK}^{ORG}\approx-0.761$.  In contrast, the BL energies split into
even and odd values, each located apparently on a simple line. Each
line separately extrapolates very close to the exact value:
$E_{SK}^{even}\approx-0.763$ and
$E_{SK}^{odd}\approx-0.765$. Amazingly, the trivial value of $e_2=-1$
is very close to the linear fit for the even results. Clearly, a
function that would interpolate continuously {\it all} the BL energies
will have to be very complicated (oscillatory). But we may speculate
that its envelope for the even $e_{k+1}$ is a simple line, passing
$e_2=-1$ and the SK result:
\begin{eqnarray} 
{\cal E}_{k+1}=E_{SK}\sqrt{k+1}-\frac{2E_{SK}+\sqrt{2}}{\sqrt{k+1}}.
\label{eveneq}
\end{eqnarray}

\begin{figure}
\vskip 2.0in 
\includegraphics{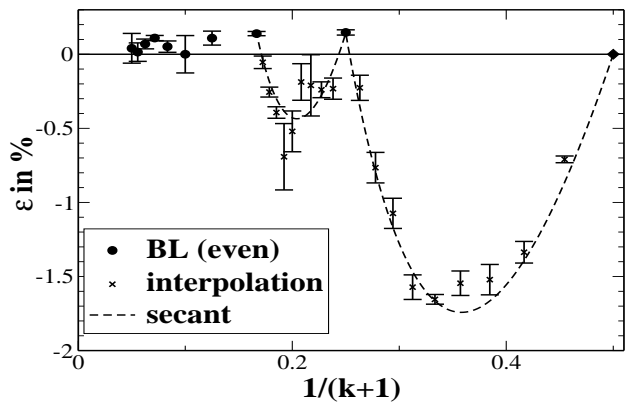}
\caption{Plot of the deviation $\epsilon$ of the BL energies for even
$k+1$ (circles) relative to Eq.~(\protect\ref{eveneq}) as a function
of $1/(k+1)$. All BL data deviates {\it at worst by 0.2\%}. The point at
$k+1=2$ (diamond) is exact. Energies from the interpolating graphs
(crosses) do not smoothly interpolate the BL data. Dashed lines
are derived from the secants $e_{k+1+2p}=pe_{k+3}+(1-p)e_{k+1}$,
$k+1=2,4$ and $0\leq p\leq1$, and clearly trace the interpolating
data.  }
\label{errorplot}
\end{figure}

In Fig.~\ref{errorplot} we plot the deviation
$\epsilon=e_{k+1}/{\cal E}_{k+1}-1$ of the extrapolated
energies from Eq.~(\ref{eveneq}) for even $k+1$. While the
extrapolated values do not fall exactly onto the proposed function,
they are {\it all} within about 0.2\% of it. In fact, all points are
slightly too high, which may indicate a more complex functional
correction to Eq.~(\ref{eveneq}), or a systematic error, say, in the
extrapolation due to higher-order corrections.

It has been pointed out \cite{Mezard_pc} that Eq.~(\ref{eveneq}) would
 imply that a first-order perturbation around the trivial $k+1=2$
 solution would be exact and give the RSB result for the SK model. But
 the obvious continuation of BL off the even integers fails to
 interpolate the data smoothly. If we ``interpolate'' BL for each
 $k+1=2,4,6,\ldots$ with a mix of $(1-p)n$ $(k+1)$-vertices and $pn$
 $(k+3)$-vertices for $0\leq p\leq1$, the resulting energies only
 provide a set of {\it secants,} $e_{k+1+2p}=pe_{k+3}+(1-p)e_{k+1}$,
 to the even-integer data. We have also plotted our (somewhat less
 accurate) extrapolation results for those interpolating graphs in
 Fig.~\ref{errorplot}. On this scale, the singular behavior of this
 continuation at the even integers becomes obvious.

We can further compare our extrapolated energies with perturbative
calculations in the SK-limit of infinite connectivity
\cite{PT,DG,LG}. A recent RSB calculation indicates a
$1/c$-correction for the ground state energy of $f_1^{BL}\approx-0.32$
for BL and $f_1^{ORG}\approx0.17$ for ORG (see Figs.~3 and~4 of
Ref.~\cite{PT}). While our (crude) fit in Fig.~\ref{SKplot} predicts
a slope at the origin of $\approx0.16$ for ORG, the slopes for either
even or odd BL data would predict $\approx0.11-0.10$, or $0.1122$ from
Eq.~(\ref{eveneq}), far from the perturbative result. It appears that
the oscillation between even and odd $k+1$ complicates also the
analytic continuation of the BL problem for large connectivities.

It will be most interesting to see how well upcoming RSB calculations
at $T=0$ for even $k+1$ will correspond to the proposed function in
Eq.~(\ref{eveneq}), or to the extrapolated energies in
Tabs.~\ref{enertable} in general. While the EO algorithm in itself can
not provide information about the physics at $T>0$, the results
presented in this Letter are sufficiently promising to apply EO also
to sample other models \cite{eo_prl} and more complicated properties
of the ground states, such as overlap distributions and excitations.

\end{document}